\pdfoutput=1
\documentclass{sf2a-conf2018}
\usepackage{graphicx}
\usepackage{hyperref}
\usepackage[]{natbib}  
\usepackage{epstopdf}

\def\BibTeX{{\rm B\kern-.05em{\sc i\kern-.025em b}\kern-.08em
    T\kern-.1667em\lower.7ex\hbox{E}\kern-.125emX}}
\bibpunct{(}{)}{;}{a}{}{,}  

\begin{document}

\TitreGlobal{The European Extremely Large Telescope and its instrument suites}


\title{Overview of the European Extremely Large Telescope and its instrument suite}

\runningtitle{The ELT ans instruments}

\author{B. Neichel}\address{LAM, Aix Marseille Univ, CNRS, CNES, Marseille, France}
\author{D. Mouillet}\address{IPAG, CNRS, Université Grenoble-Alpes, 38041 Grenoble, France}
\author{E. Gendron}\address{LESIA, Observatoire de Paris, Universite PSL, CNRS, Sorbonne Université, Univ. Paris Diderot, Sorbonne Paris Cite, 5 place Jules
Janssen, 92195 Meudon, France}
\author{C. Correia$^1$}
\author{J.\,F. Sauvage$^{1,}$}\address{ONERA, The French Aerospace Lab BP72, 29 avenue de la Division Leclerc, 92322 Chatillon Cedex, France} 
\author{T. Fusco$^{1,4}$}




\setcounter{page}{237}


\maketitle


\begin{abstract}
The European Extremely Large Telescope will see first lights by the end of 2024. With a diameter of almost 40 meters, it will be the biggest optical telescope ever built from the ground. The ELT will open a brand new window in a sensitivity / spatial angular resolution parameter space. To take the full benefit of the scientific potential of this giant, all the instruments will be equipped with Adaptive Optics (AO), providing the sharpest images. This paper provides a quick overview of the AO capabilities of the future instruments to be deployed at the ELT, highlight some of the expected performance and describe a couple of technical challenges that are still to tackle for an optimal scientific use. This paper has been presented at the "{\it Societe Francaise d'Astronomie et Astrophysique}" symposium in Bordeaux 2018, it is then naturally biased toward the French contribution for the ELT.
\end{abstract}



\section{Introduction}
 Since the first telescopes built more than 400 years ago, ground-based astronomy has been driven by the quest for larger aperture, higher angular resolution, larger Field-of-View (FoV) and larger wavelength coverage. After the scientific success of 8/10 meter telescopes like the VLT, Keck, Gemini, Subaru, etc... international teams have been putting effort in designing the next generation of large ground based telescope: the generation of the Extremely Large Telescopes (ELT). There are two obvious reasons why astronomers are interested in building larger apertures. First is the gain in collecting power, that scales as the surface of the primary mirror. As such, when going from an 8m telescope to a 40m telescope, the effective area is multiplied by a factor 25. A first way to figure out this gain is to think that what will be possible with 1 night of ELT, will require 25 nights of VLT. The telescope efficiency is then extremely increased. Another way to realize the step in terms of collecting power, is to imagine that the ELT surface will be larger than all the 8/10m telescopes currently in operations.\\
 The other main reason to build larger and larger telescope comes from the gain in theoretical angular resolution. The resolving power of a telescope (or its ability to distinguish details) is directly proportional to the diameter of its aperture. From a VLT to an ELT, the resolving power will be improved by a factor 5. However, this last gain is true, if and only if, the telescope works at its diffraction limit, which means that all the optical aberrations induced by the telescope itself, or by the atmospheric turbulence are corrected. This is the purpose of Adaptive Optics (AO), and all the ELTs will be equipped with dedicated AO instruments, to take the full benefit of the large apertures. In this paper we will first describe one of the ELT - the European one led by ESO, then we will give a quick description of its instrument suite, focusing mostly on the AO capabilities of each. Finally, we will highlight a couple of the main challenges that are to be solved for the first light of those giant telescopes. 
 \begin{figure}[ht!]
 \centering
 \includegraphics[width=0.8\textwidth,clip]{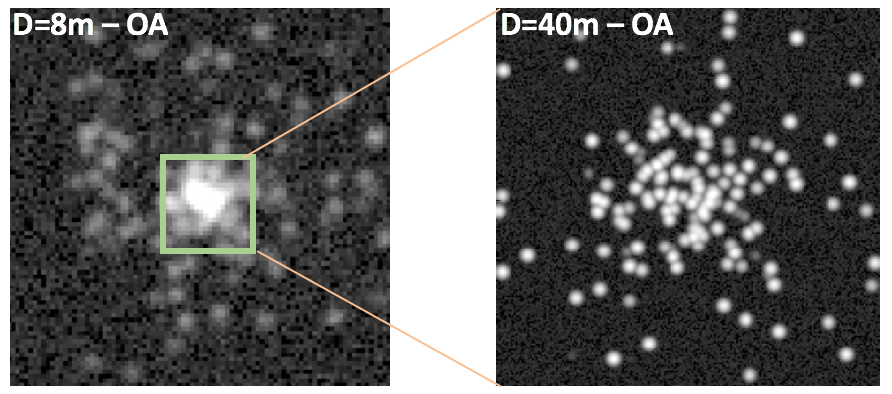}      
  \caption{Illustration (simulation) of the gain brought by a larger telescope aperture, when going from 8m to 40m. The gain is both in collecting power, and in angular resolution. The total gain in sensibility (capability to detect faint objects) scales as the diameter to power of 4.}
  \label{author1:fig1}
\end{figure}
  
\section{The European Extremely Large Telescope}
There are three large projects to build ELTs currently on-going, namely the Thirty Meter Telescope (TMT - \url{www.tmt.org}), the Giant Magellan Telescope (GMT - \url{www.gmto.org}) and the Extremely Large Telescope (ELT - \url{https://www.eso.org/sci/facilities/eelt/}). All three are adopting different strategies in terms of telescope design, and we will focus here on the European one - the ELT. The ELT implements an innovative optical design, diverging from the classical Gregorian or Ritchey-Chrétien 3-mirrors designs. Indeed, the ELT is a 5 mirrors telescope, which ensure both a wide field (10 arcminutes) and a good optical quality (basically diffraction limited over the whole field). \\
This primary mirror is a giant segmented mirror, made of 798 hexagonal segments of 1.4m each. All the segments are controlled in positions for Tip-Tilt and piston, in order to ensure the telescope phasing. This active control is also required to counter the gravity effects, and wind-induced deformations. On top of those degrees of freedom, the specificity of the ELT is to include a deformable mirror in its optical train: the fourth mirror (a.k.a. M4 - \cite{Vernet12}) has almost 6000 actuators that can be controlled at high temporal speed (up to 1000Hz). Those actuators are required to maintain the optical quality of the telescope, and they are continuously used during operations \citep{Bonnet18}. The ELT is too large to operate in conventional seeing-limited mode, and an adaptive correction is needed all the time. But the level of correction provided by the telescope, only ensures that the image quality provided is seeing-limited. The telescope specification being that: "The telescope shall deliver seeing-limited performance with natural guide star(s), and not degrade the FWHM of a point source generated by an ideal telescope operated in the atmosphere by more than 5\%". This optical quality should be delivered for most of the seeing conditions. To reach this performance, and control the degrees of freedom provided by M4, the telescope is equipped with several Wave-Front Sensors (WFS) in a Pre-Focal Station (PFS - \cite{Lewis18}). \\
The next level of correction, in order to bring the images at the diffraction limit is provided by the instruments. The ELT-instrument documents describe that "Enhancement of the image quality beyond the seeing limited performance shall be achieved via a combination of the telescope, instruments and adaptive optics modules working together". It is then responsibility of the instruments, hence the community, to improve the angular resolution of the ELT by the remaining factor 100.

 \begin{figure}[ht!]
 \centering
 \includegraphics[width=0.9\textwidth,clip]{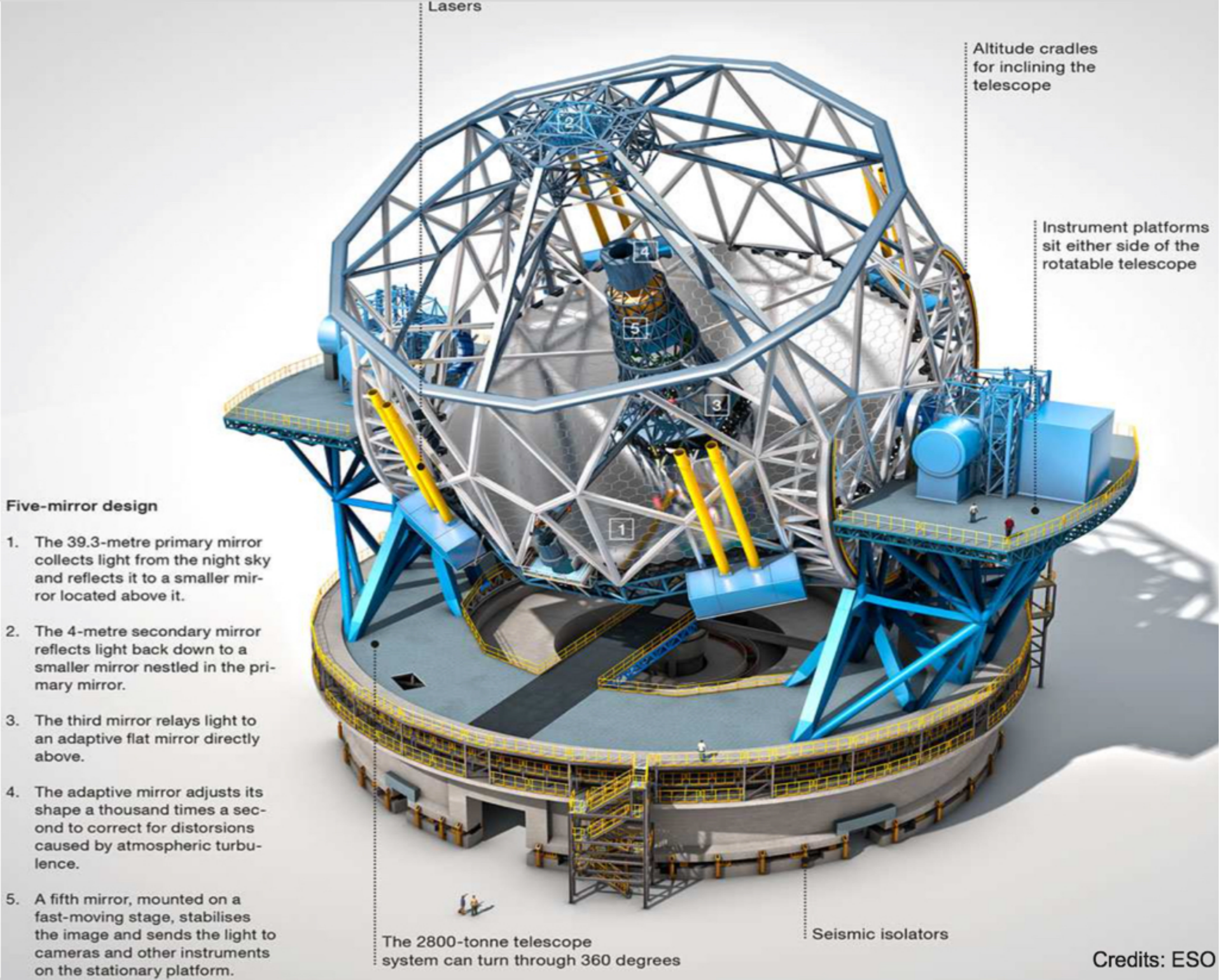}      
  \caption{View of the ELT, with details on each of its 5 mirrors. Credits ESO.}
  \label{author1:fig1}
\end{figure}

\section{The ELT instruments}

\begin{figure}[ht!]
 \centering
 \includegraphics[width=0.9\textwidth,clip]{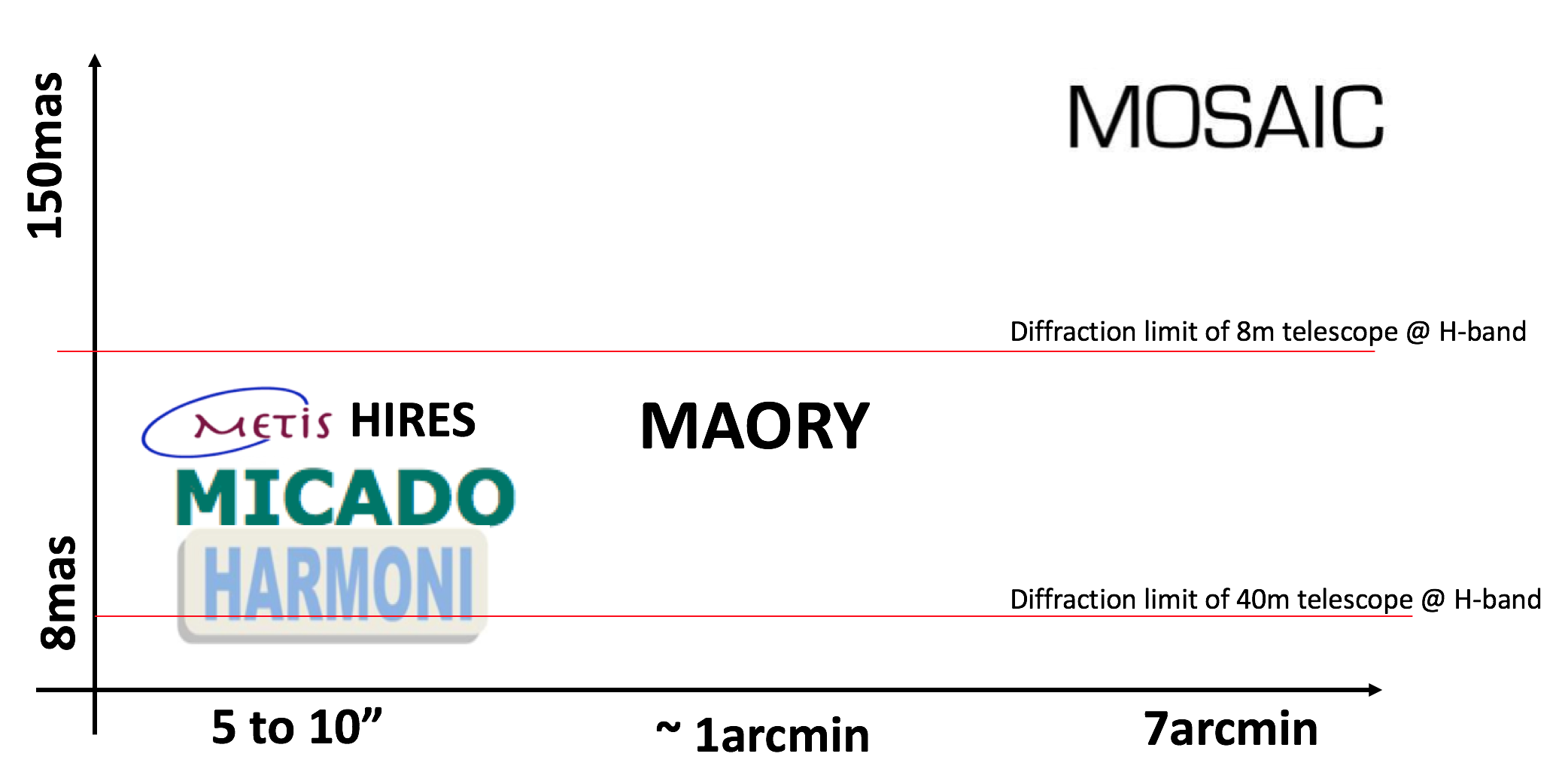}      
  \caption{Overview of the ELT instrument suite, in an angular resolution vs. field of view diagram. The diffraction limit of the ELT (8mas at H-band), and the diffraction limit of the VLT (40mas) are shown for reference.}
  \label{instrum1b}
\end{figure}

The instrument roadmap of the ELT has been described as early as 2006 in the following document: \url{https://www.eso.org/sci/facilities/eelt/docs/E-ELT-Construction-Proposal-INS-Chapter.pdf}. It included two first light instruments: an imager (ELT-CAM) and an IFU (ELT-IFU), and a first generation mid-infrared instrument (ELT-MIR). The development plan then considered second-generation instruments with a MOS (ELT-MOS) and a high spectral resolution (ELT-HIRES). After the Phase-A studies, and delta Phase-A studies, those instruments became respectively: MICADO/MAORY (ELT-CAM - \cite{Davies18, Ciliegi18}), HARMONI (ELT-IFU - \cite{Thatte16, Neichel16}), METIS (ELT-MIR - \cite{Brandl18}), MOSAIC (ELT-MOS - \cite{Morris18}) and HIRES (ELT-HIRES - \cite{Marconi18}). As described above, it is the responsibility of each instrument to develop its own AO strategy, taking benefit of the adaptive telescope, and each one could pursue different AO flavor, depending on its science cases \citep{Ramsay18}. A first way to represent all the different AO performance is shown in Fig. \ref{instrum1b}, which shows the performance (in terms of angular resolution) vs. the Field of View (FoV) accessible by each instrument. It is first interesting to note that most of the first light instruments are exploiting the highest angular resolution of the ELT - which is the obvious scientific niche. The trade-off being made on the accessible FoV. Indeed, with a fixed money cost for the instrument development, the number of pixels that each instrument can afford is limited. A trade-off has to be made between FoV and pixel size. For MOSAIC, the science case pushes for larger FoV, and the angular resolution is decreased. But even though the resolution is lower than on could reach with a diffraction-limited instrument on a 8m telescope, the instrument benefits from the collecting power of the 40m ELT aperture.  

\begin{figure}[ht!]
 \centering
 \fbox{\includegraphics[width=0.5\textwidth,clip]{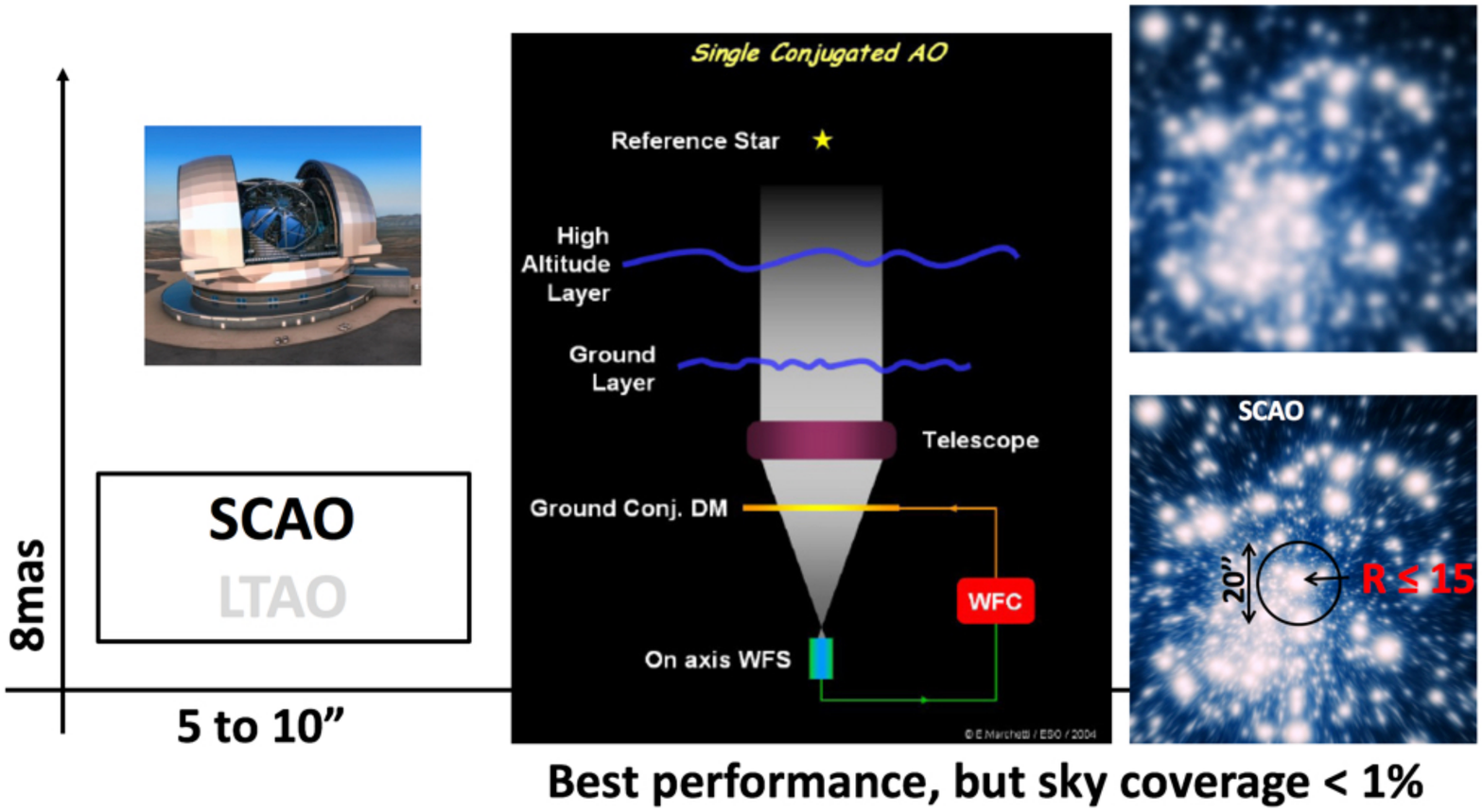}}%
 \fbox{\includegraphics[width=0.5\textwidth,clip]{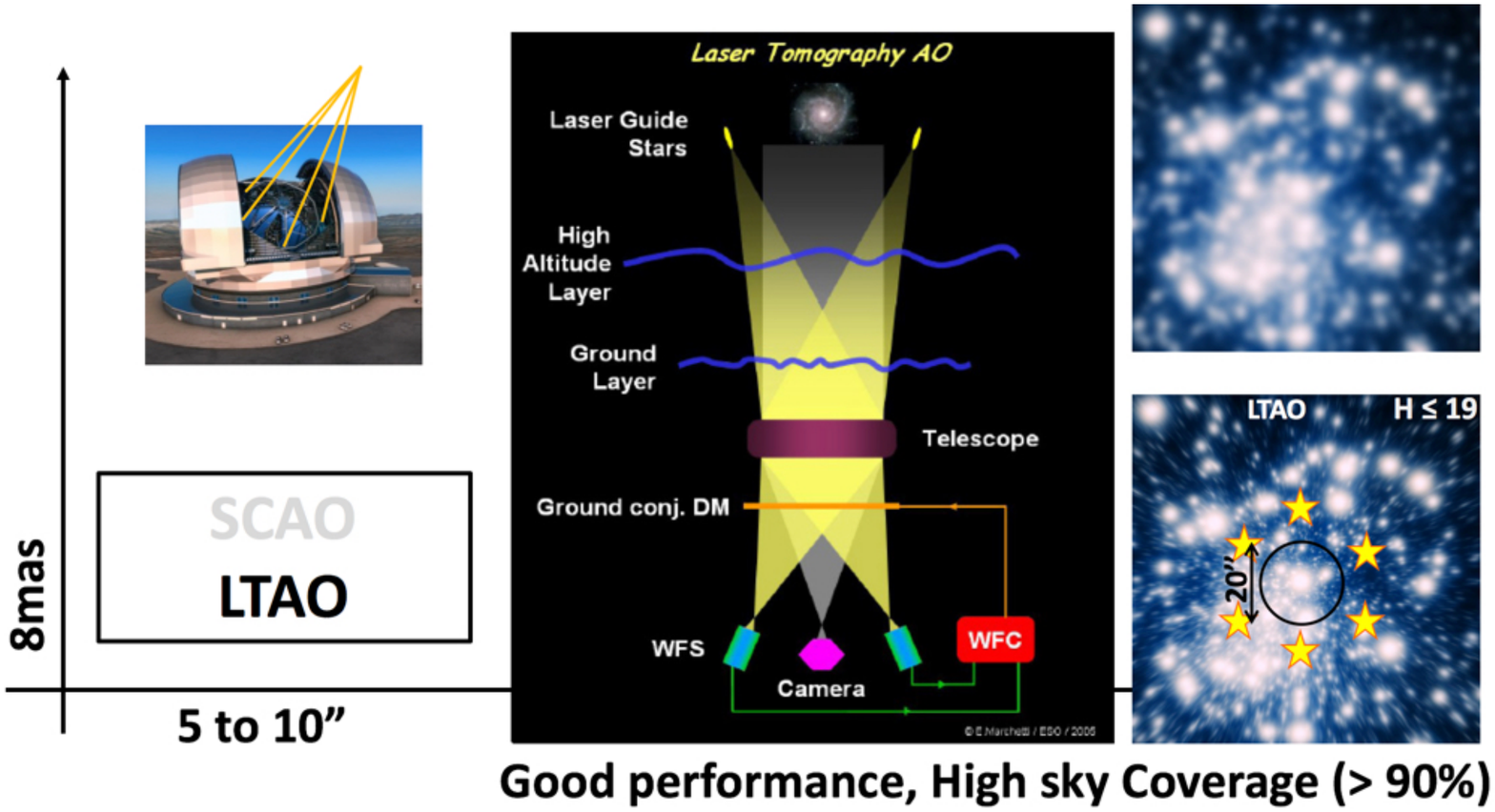}} \\
  \fbox{\includegraphics[width=0.5\textwidth,clip]{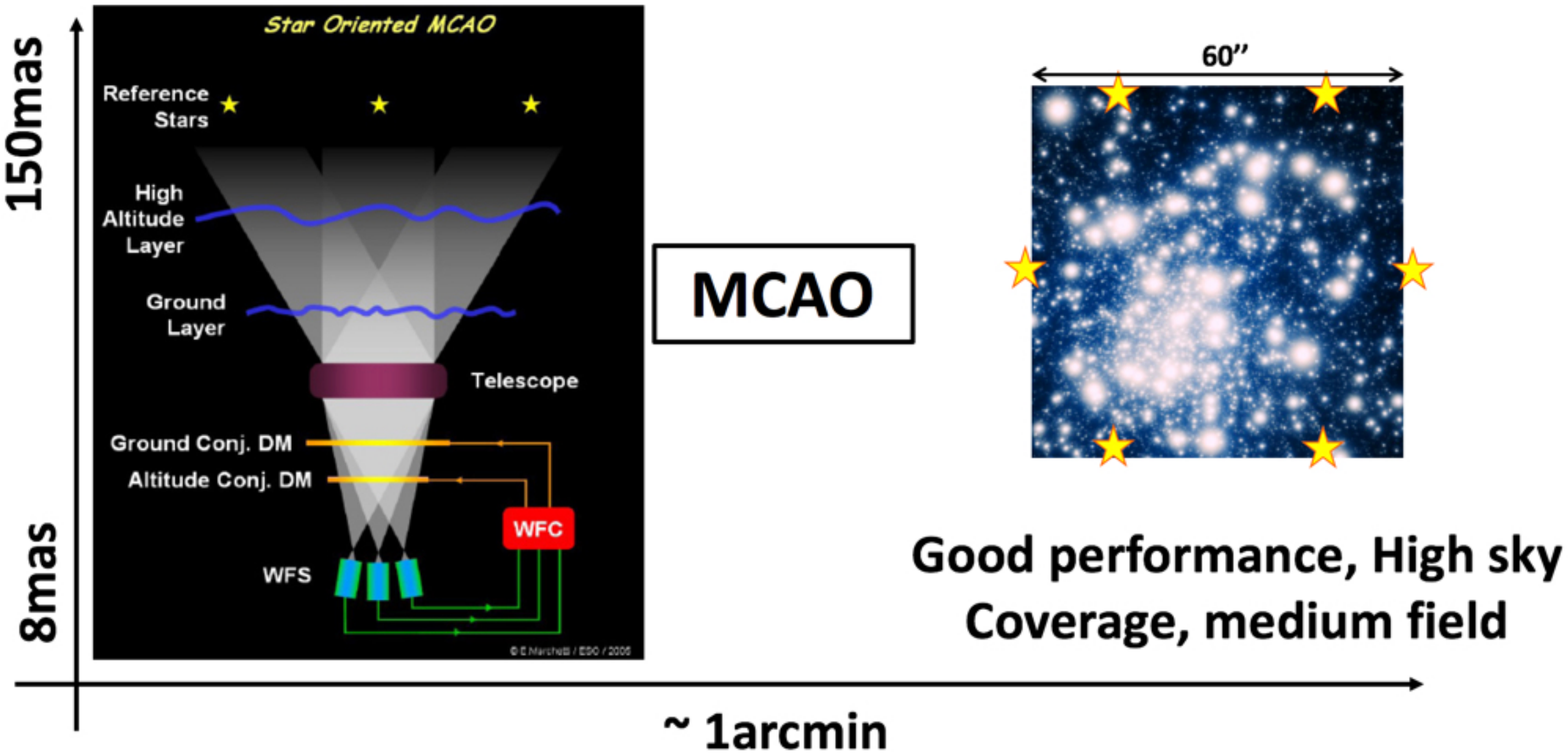}}%
 \fbox{\includegraphics[width=0.5\textwidth,clip]{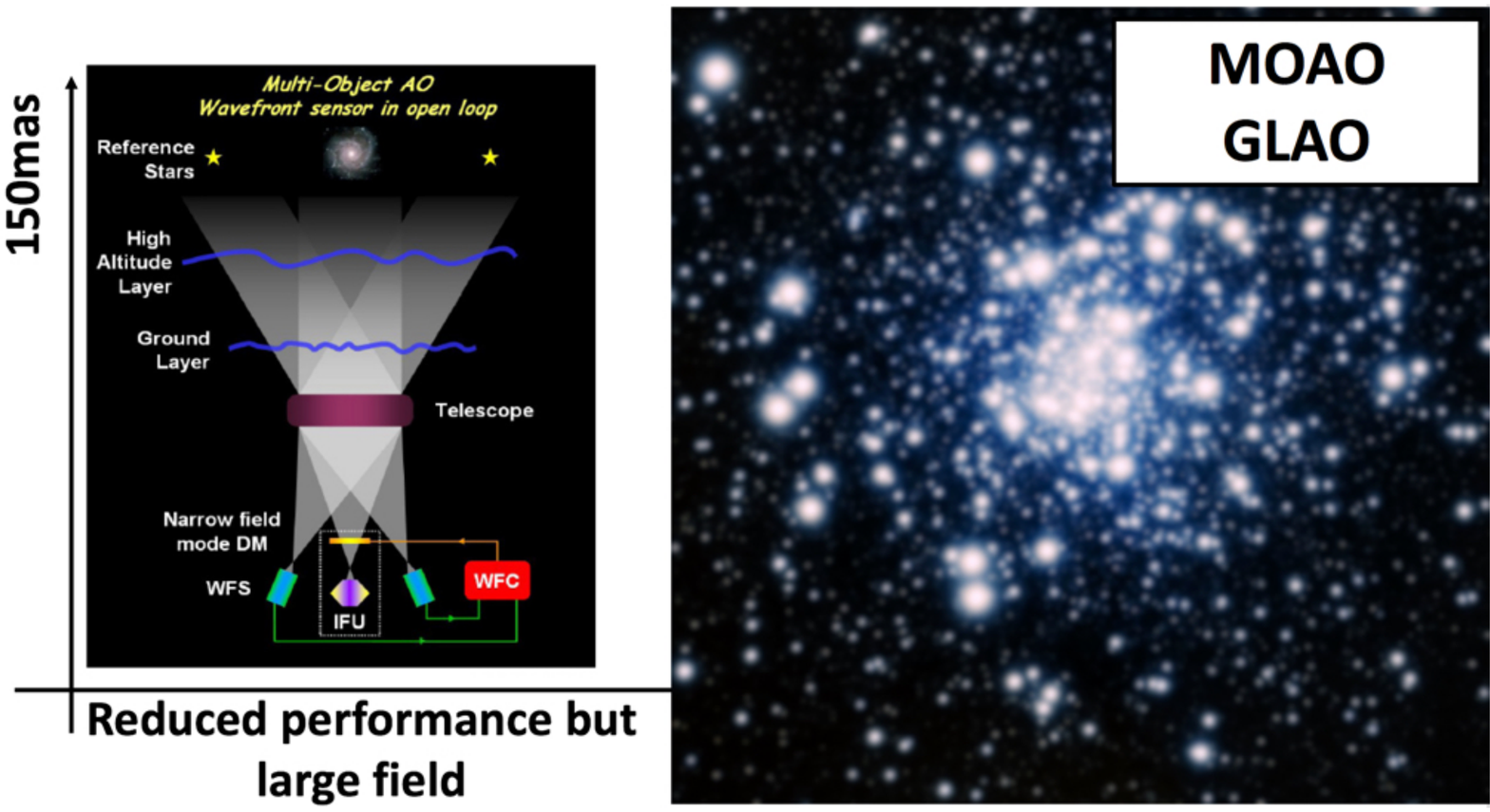}}       
  \caption{Illustration of the different AO flavors to be implemented for the ELT. {\bf Top-Left:} SCAO will be implemented by HARMONI, MOSAIC, MICADO and HIRES. {\bf Top-Right:} LTAO will be implemented by HARMONI {\bf Bottom-Left:} MCAO will be implemented by MAORY to feed MICADO. {\bf Bottom-Right:} GLAO and MOAO will be implemented by MOSAIC.}
  \label{ao}
\end{figure}

Different spatial angular performance are reached with different implementation of the AO modules. Figure \ref{ao} shows the different AO flavors that will be implemented on the ELT. 
\begin{itemize}
\item {\bf SCAO:} stands for Single Conjugate Adaptive Optics. This is the most basic AO implementation, where a single Wave-Front Sensor (WFS) is used on Natural Guide Star (NGS). SCAO provides the best performance, brings the images to the diffraction-limit of the 40m telescope, but requires bright and close enough reference stars. Typically, a NGS with a magnitude higher than R=14, and within a radius of $\sim$15arcsec should be used. As a consequence, the associated sky coverage (or equivalently the probability than a given target may be observed) is extremely low, less than a percent. HARMONI, MICADO and HIRES will implement SCAO systems \citep{Clenet18, Xompero18, Neichel16}.
\item {\bf LTAO:} stands for Laser Tomography Adaptive Optics. In order to tackle the sky-coverage issue of SCAO, it is possible to use artificial reference star, a.k.a. Laser Guide Stars (LGSs). In that case, a very bright laser is propagated from the telescope, up to a high altitude layer of Sodium atoms located at $\sim$90km above the telescope. The light from the LGS, exactly shining at 589nm, is absorbed and re-emitted by the sodium atoms, creating an artificial light source that may be used by the WFSs. Because the LGSs are at a finite altitude, they do not sense the same volume of atmosphere as the one coming the scientific target. Hence, it is required to use several LGSs in parallel, and combine their signals in a tomographic way. The ELT will implement between 4 and 8 LGSs that can be used by the instruments. An LTAO system then provides almost similar performance as a SCAO system, however, over a fraction of the sky which is now almost complete. The sky coverage is not 100\%, because at least one NGS is still required to compensate for some corrupted measurements of the LGSs. But this NGS may be fainter (typically H$<$19), and could be picked at a largest distance from the scientific target (typically 1 arcmin). HARMONI will implement an LTAO system \citep{Neichel16}.
\item {\bf MCAO:} stands for Multi-Conjugate Adaptive Optics. An LTAO system solves the sky-coverage limitation, but the correction provided is only optimized for a small FoV (typically less than 10 arcsec). If one wants to increase the corrected FoV, more degrees of freedom are required. This is achieved by implementing post-focal deformable mirrors, that are used in conjunction with M4. With more deformable mirrors, and the several LGSs, an MCAO system can deliver diffraction limited performance over a field that can reach 1 or 2 arcminutes. MAORY is the MCAO module of the ELT \citep{Diolaiti16}. It will feed MICADO.
\item {\bf GLAO:} stands for Ground-Layer Adaptive Optics. If one wants to significantly increase the corrected FoV, trade-offs are to be made on the level of correction provided by the AO system. Indeed, a fully corrected 5 or 7 arcminutes field would required more than 5 post-focal DMs in an MCAO configuration, which makes such an instrument out of scope of complexity and cost. With a single deformable mirror, as is M4, but combining WFSs measurements from far off-axis LGSs or NGSs, the level of correction provided by a GLAO system will be partial, but uniform of a large field. A GLAO system only compensates for the atmospheric turbulence in the first hundred of meters above the telescope, but those are usually the most energetic ones. As such, a GLAO correction will not provide diffraction limited images, but typically shrink the seeing PSF image by a factor 2 to 5. It can be seen as seeing-reducer, shifting the median seeing of Armazones from $\sim$0.65 arcsec down to $\sim$0.2 or 0.3arcsec. MOSAIC intends to use a GLAO correction for its High-Multiplex Mode (HMM - \cite{Morris18b}).
\item {\bf MOAO}: stands for Multi-Object Adaptive Optics. One way to improve the performance over a very large FoV is to provide local corrections, with dedicated deformable mirrors. MOAO systems are mostly driven by extra-galactic science cases, where it is not needed to have a full corrected FoV, but only focus on specific directions: where the galaxies are. As a second stage after the telescope M4, small DMs can be adjusted on the FoV to provide such a dedicated correction. MOSAIC implements an MOAO correction for its High-Definition Mode (HDM - \cite{Morris18b}). 
\end{itemize}

The French teams are deeply involved in the development of the different AO modes of the instruments. In particular, France is in charge of the SCAO system for MICADO, the SCAO system for HARMONI, the LTAO system for HARMONI, the development of the LGS WFS for MAORY and has a shared responsibility in the development of the GLAO and MOAO modes of MOSAIC. This is a major investment of the French AO community, and a long-term effort that started more than 10 years ago, and will last for the next decade (at least).

\section{Some challenges}
\begin{figure}[ht!]
 \centering
 \includegraphics[width=0.99\textwidth,clip]{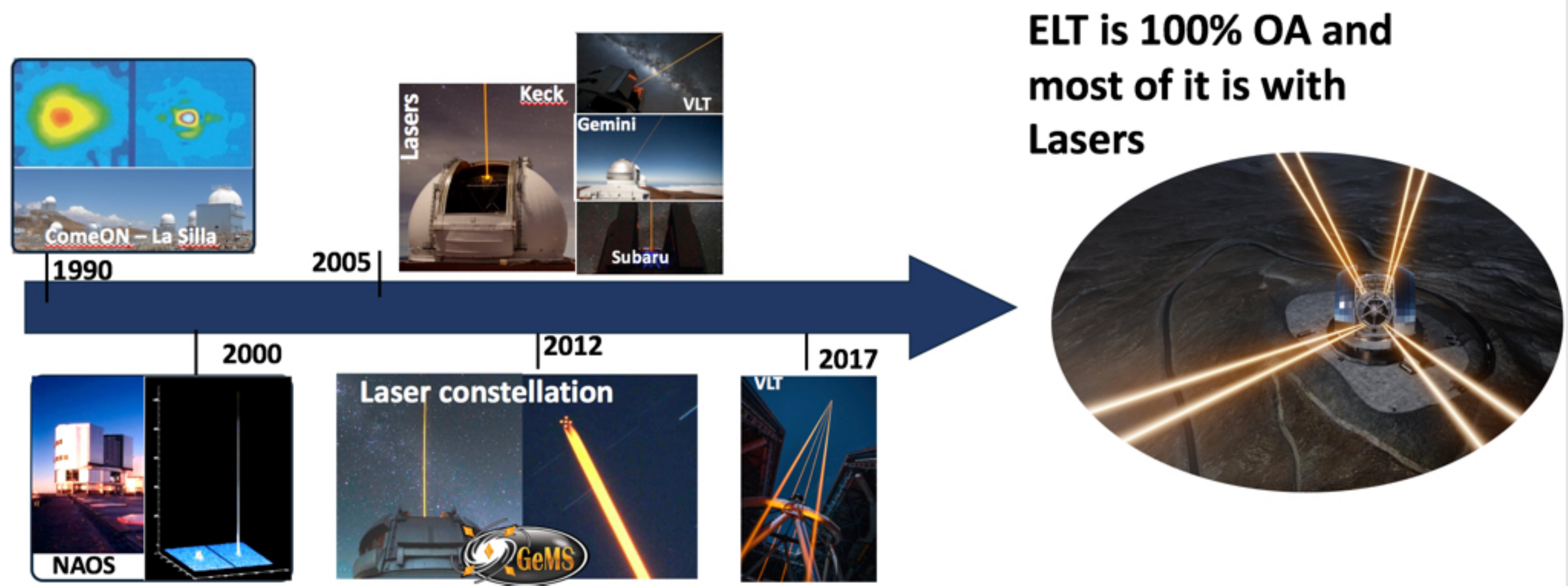}      
  \caption{(Personal) Overview of some of the main milestone reached in Adaptive Optics instruments since the last 30years. The first use of LGS date from 2005, and multiple LGS only since 2012, while it will become the standard for the ELT.}
  \label{schedule2}
\end{figure}

Building an AO system for a 40m telescope is, by nature, challenging. Even though the community has been building ambitious AO systems over the past 15 years, like SPHERE for instance \citep{Sauvage16}, the change of scale when going to an ELT poses several challenges. One of the first obvious illustration, is to realize that while the angular resolution is shrinked by a factor 5 with respect to an 8m telescope, the size of the telescope, and therefore all the vibrations, wind-load and other environmental effects are increased by a factor much larger. The size of the whole telescope structure becomes gigantic, while we have to control the optical axis stability to few microns \citep{Bonnet18}. \\
Another challenge, due to scaling factors, comes from the real-time control of M4, and other deformable mirrors. The real-time machines will have to deal with $\sim$50000 measurements, to control $>$6000 actuators, at a frame rate of almost 1kHz. This implies Tera floating-point operations per second, and dedicated hardware and software has to be developed \citep{Dunn18, Gratadour18}.\\
In terms of WFSensing, which is under the responsibility of the community, two main challenges are identified. First the SCAO systems are using a novel type of WFS, called pyramid WFS. The concept has been introduced in 1996 by Ragazzoni \citep{Ragazzoni96}, however, the operational use of such sensors is fairly new (2012 at LBT - \cite{Esposito12}), and most importantly, none of the AO systems installed at the VLT are implementing such sensors. All the teams selected this WFS because it provides better performance (better sensitivity), but there is a lack of operational experience that the French community is addressing with several lab experiments and on-sky demonstrations \citep{Bond16, Deo18}. \\
Another challenge comes from the use the LGSs on an ELT. Indeed, and as mentioned above, these artificial stars are not perfect, an in particular they suffer from what is called "elongation". A LGS source is not a punctual object in the sky, but as the Sodium layer has an intrinsic width of $\sim$10 - 15km, the LGS source appears like a cigar. When seen from the side, the perspective produces an elongated object on the WFSs. This elongation may be as large as 20arcsec, inducing biases in the measurements. Again, a large effort has been devoted in France to better understand this effect, and develop robust strategies for on-sky operations. Both with tests at the William Hershel telescope, and intense realistic simulations \citep{Bardou18, Neichel16, Oberti18}.\\ 
From a more conceptual point of view, one could also ask for the maturity of the different AO concepts. Indeed, and as shown above, the ELT will be 100\% AO, and most of it being with multiple LGSs systems. As of today, only 2 instruments are implementing multi-LGS operations \citep{Rigaut14, Neichel14, }, as shown by Fig. \ref{schedule2}. The recently commissioned AOF, feeding MUSE both in LTAO and GLAO has provided very useful insights from the operational point of view, and can clearly be identified as a pathfinder for the ELT.

\begin{figure}[ht!]
 \centering
 \includegraphics[width=0.5\textwidth,clip]{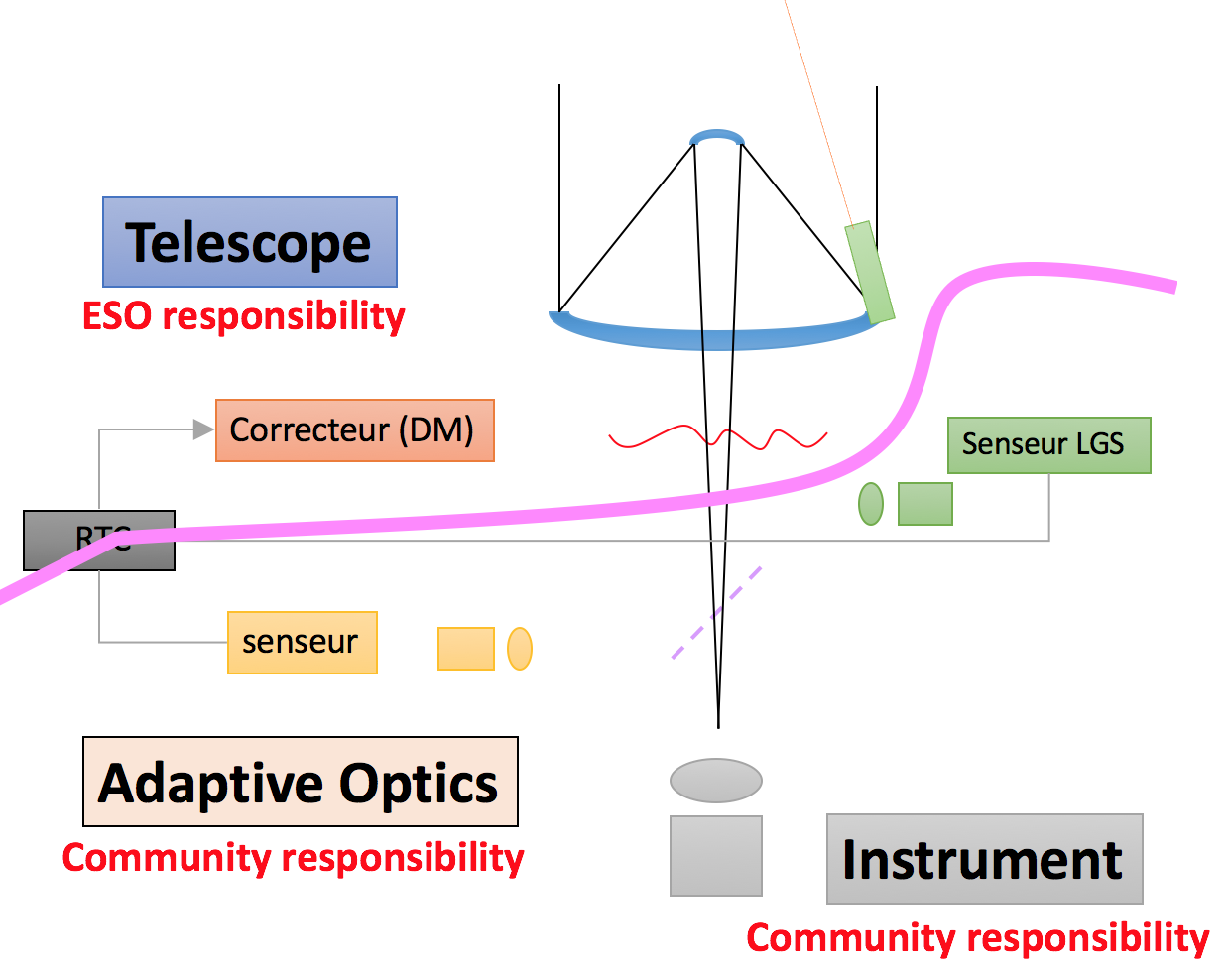}      
  \caption{Schematic, and simplified view, of the shared responsibility between the Telescope (ESO) and the Instruments (community) for the AO systems.}
  \label{resp2}
\end{figure}

\noindent Finally, from a programatic point of view, one of the big challenge imposed by the ELT concerns the interfaces between the telescope, and the instruments. In previous AO instruments, there were a clear cut between the telescope and the post-focal AO systems. The telescope management was under ESO responsibility, while the WFS, Deformable mirrors and Real-Time controllers were under the instrument responsibility. Even when using LGSs, the split was clear, with the telescope providing the LGSs sources, and the instruments providing the LGSWFSs. With the ELT, the situation is different, as the deformable mirror is now part of the telescope, as is part of the real-time control. This is somehow illustrated in Fig. \ref{resp2}. Both the telescope Pre-Focal Station WFSs, and the instrument WFSs are controlling M4, hence a clear definition of the interfaces is required, and the final success will only be possible if the telescope control and the instruments work together. 

\section{Conclusions}
The ELT will be the biggest ground-based telescope ever built. It will start operation around 2025, and will be equipped with a full suit of instruments. The science cases of the ELT and its instruments fully rely on the performance of Adaptive Optics systems. The colossal size of the telescopes (39 m) and the complexity of the scientific instruments compel us on a complete rethinking, in order to improve the overall performance, but more specifically the sensitivity and the robustness of the AO systems, and thus to maximize the astrophysical returns of AO assisted instruments. AO is at the heart of the ELT, and a large effort has been devoted in the community, and in particular in France, to address all the technical challenges raised. 

\bibliographystyle{aa}  

\begin{thebibliography}{0}
\expandafter\ifx\csname natexlab\endcsname\relax\def\natexlab#1{#1}\fi

\end{thebibliography}


\begin{thebibliography}{}
\bibitem[Bardou et al. (2018)]{Bardou18} Bardou, L., Gendron, E., Rousset, G. et al. 2018, SPIE, Volume 10703, id. 107031X
\bibitem[Bond et al. (2016)]{Bond16} Bond C., El Hadi, K., Sauvage, J.-F. et al. 2016, SPIE, Volume 9909, id. 990964
\bibitem[Bonnet et al. (2018)]{Bonnet18} Bonnet, H., Biancat-Marchet, F., Dimmler, M. et al. 2018, SPIE, Volume 10703, id. 1070310
\bibitem[Brandl et al. (2018)]{Brandl18} Brandl, B. R., Absil, O., Agocs, T. et al. 2018, SPIE, Volume 10702, id. 107021U
\bibitem[Ciliegi et al. (2018)]{Ciliegi18}Ciliegi, P., Diolaiti, E., Abicca, R. et al. 2018, SPIE, Volume 10703, id. 1070311
\bibitem[Clenet et al. (2018)]{Clenet18} Clenet, Y., Buey, T., Gendron, E. et al. 2018, SPIE, Volume 10703, id. 1070313
\bibitem[Davies et al. (2018)]{Davies18} Davies, R., Alves, J., Clénet, Y et al. 2018, SPIE, Volume 10702, id. 107021S 
\bibitem[Deo et al. (2018)]{Deo18} Deo V., Gendron E., Rousset E. et al. 2018, SPIE, Volume 10703, id. 1070320
\bibitem[Dunn et al. (2018)]{Dunn18} Dunn, J., Kerley, D., Smith, M. et al. 2018, SPIE, Volume 10703, id. 1070317
\bibitem[Diolaiti et al. (2016)]{Diolaiti16} Diolaiti, E., Ciliegi, P., Abicca, R. et al. 2016, SPIE, Volume 9909, id. 99092D
\bibitem[Esposito et al. (2012)]{Esposito12} Esposito, S., Riccardi, A., Pinna, E. et al. 2012, SPIE, Volume 8447, article id. 84470U
\bibitem[Gratadour et al. (2018)]{Gratadour18} Gratadour, D., Morris, T., Biasi, R. et al 2018, SPIE, Volume 10703, id. 1070318
\bibitem[Morris et al. (2018)]{Morris18} Morris, S., Hammer, F., Jagourel P. et al. 2018, SPIE, Volume 10702, id. 107021W
\bibitem[Morris et al. (2018)]{Morris18b} Morris, T., Basden, A., Calcines-Rosario, A. et al., 2018, SPIE, Volume 10703, id. 1070316
\bibitem[Lewis et al. (2018)]{Lewis18} Lewis, S. A. E., Brunetto, E., Forster, A. et al. 2018, SPIE, Volume 10700, id. 107001B
\bibitem[Marconi et al. (2018)]{Marconi18} Marconi, A., Allende Prieto, C., Amado, P. J et al. 2018, SPIE, Volume 10702, id. 107021Y
\bibitem[Neichel et al. (2014)]{Neichel14} Neichel, B., Rigaut, F., Vidal, F. et al. 2014 MNRAS Volume 440, Issue 2, p.1002-1019
\bibitem[Neichel et al. (2016)]{Neichel16} Neichel, B., Fusco, T., Sauvage, J.F. et al. 2016, SPIE, Volume 9909, id. 990909
\bibitem[Oberti et al. (2018)]{Oberti18} Oberti, S., Kolb, J., Madec, P.-Y. et al. 2018, SPIE, Volume 10703, id. 107031G
\bibitem[Ragazzoni (1996)]{Ragazzoni96} Ragazzoni, R., 1996, Journal of Modern Optics, vol. 43, Issue 2, p.289-293
\bibitem[Ramsay et al. (2018)]{Ramsay18} Ramsay, S., Casali, M., Amico, P. et al. 2018 SPIE, Volume 10702, id. 107021P
\bibitem[Rigaut at al. (2014)]{Rigaut14} Rigaut, F., Neichel, B., Boccas, M. et al. 2014 MNRAS, Volume 437, Issue 3, p.2361-2375
\bibitem[Sauvage et al. (2016)]{Sauvage16} Sauvage, J.-F., Fusco, T., Petit, C. et al. 2016, JATIS, Volume 2, id. 025003
\bibitem[Thatte et al. (2016)]{Thatte16} Thatte, N., Clarke, F., Bryson, I. et al. 2016, SPIE, Volume 9908, id. 99081X
\bibitem[Vernet et al. (2012)]{Vernet12} Vernet, E., Cayrel, M., Hubin, N. et al., 2012, SPIE, 8447, Adaptive Optics Systems III, 844761
\bibitem[Xompero et al. (2018)]{Xompero18} Xompero, M., Giordano, C., Bonaglia, M. et al. 2018, SPIE, Volume 10703, id. 1070341
\end{thebibliography}

%
\end{document}